\documentclass[twocolumn,english,aps,pra,showpacs]{revtex4-1}
\usepackage[T1]{fontenc}
\usepackage[latin1]{inputenc}
\usepackage{amsmath}
\usepackage{graphicx}
\usepackage{amssymb}
\usepackage{esint}

\makeatletter

\providecommand{\tabularnewline}{\\}

\@ifundefined{textcolor}{}
{%
 \definecolor{BLACK}{gray}{0}
 \definecolor{WHITE}{gray}{1}
 \definecolor{RED}{rgb}{1,0,0}
 \definecolor{GREEN}{rgb}{0,1,0}
 \definecolor{BLUE}{rgb}{0,0,1}
 \definecolor{CYAN}{cmyk}{1,0,0,0}
 \definecolor{MAGENTA}{cmyk}{0,1,0,0}
 \definecolor{YELLOW}{cmyk}{0,0,1,0}
 }

\makeatother

\makeatother

\usepackage{babel}

\makeatother

\usepackage{babel}

\begin{document}

\title{Universal contact of strongly interacting fermions at finite temperatures}

\author{Hui Hu$^{1}$}

\email{hhu@swin.edu.au}

\affiliation{$^{1}$ARC Centre of Excellence for Quantum-Atom Optics, Centre for
Atom Optics and Ultrafast Spectroscopy, Swinburne University of Technology,
Melbourne 3122, Australia}

\author{Xia-Ji Liu$^{1}$}

\email{xiajiliu@swin.edu.au}

\affiliation{$^{1}$ARC Centre of Excellence for Quantum-Atom Optics, Centre for
Atom Optics and Ultrafast Spectroscopy, Swinburne University of Technology,
Melbourne 3122, Australia}

\author{Peter D. Drummond$^{1}$}

\email{pdrummond@swin.edu.au}

\affiliation{$^{1}$ARC Centre of Excellence for Quantum-Atom Optics, Centre for
Atom Optics and Ultrafast Spectroscopy, Swinburne University of Technology,
Melbourne 3122, Australia}

\date{\today{}}
\begin{abstract}
The recently discovered universal thermodynamic behaviour of dilute,
strongly interacting Fermi gases also implies a universal structure
in the many-body pair-correlation function at short distances, as
quantified by the contact ${\cal I}$. Here we theoretically calculate
the temperature dependence of this universal contact for a Fermi gas
in free space and in a harmonic trap. At high temperatures above the
Fermi degeneracy temperature, $T\gtrsim T_{F}$, we obtain a reliable
non-perturbative quantum virial expansion up to third order. At low
temperatures we compare different approximate strong coupling theories.
These make different predictions, which need to be tested either by
future experiments or advanced quantum Monte Carlo simulations. We
conjecture that in the universal unitarity limit, the contact or correlation
decreases monotonically with increasing temperature, unless the temperature
is significantly lower than the critical temperature, $T\ll T_{c}\sim0.2T_{F}$.
We also discuss briefly how to measure the universal contact either
in homogeneous or harmonically trapped Fermi gases. 
\end{abstract}

\pacs{03.75.Hh, 03.75.Ss, 05.30.Fk}

\maketitle

\section{Introduction}

Understanding strongly interacting fermions is one of the most challenging
problems in present-day physics \cite{ohara}. While the behavior
of interacting fermions can be understood in some well-defined regions
of parameter space, it remains elusive in the correlated strongly
interacting regime \cite{rmpgiorgini,rmpbloch}. A better understanding
of strongly interacting fermions has wide-ranging implications for
systems such as quark matter in neutron stars and high-temperature
superconductors. An important generic idea in this field is fermionic
universality \cite{houniversality,hdlnatphys}: all strongly interacting,
dilute Fermi gases should behave identically, depending only on a
scaling factor equal to the average particle separation, but not on
the details of the interaction.

The recent realization of broad collisional (Feshbach) resonances
with an interaction potential range $r_{0}$ small relative to the
inter-particle spacing in ultracold atomic Fermi gases provides a
highly controlled environment for studying the general problem of
strongly interacting fermions \cite{rmpgiorgini,rmpbloch}. By applying
an external magnetic field across the resonance, the interparticle
interaction can be accurately tuned from weak to infinitely strong
\cite{rmpchin}. This has led to the observation of crossover from
Bardeen-Cooper-Schrieffer (BCS) superfluids to Bose-Einstein condensations
(BEC) \cite{nsr,jilaCrossover,mitCrossover}. 

The most strongly interacting regime lies at the resonance, where
the \textit{s}-wave scattering length $a_{s}$ diverges and the two-body
scattering amplitude reaches its maximum value allowed by quantum
mechanics, \textit{i.e.}, it becomes unitarity limited \cite{houniversality}.
A number of properties in the unitarity limit have been characterized.
In particular the universal thermodynamic behavior \cite{hdlnatphys,jilaEoS,dukeEoS1,dukeEoS2,ensEoS1,ensEoS2,tokyoEoS}
has been experimentally observed, and agrees with approximate analytic
theories and Monte Carlo simulations. However, it is still nontrivial
to calculate these universal properties quantitatively, as there is
no small interaction parameter in a perturbation theory expansion.

In 2008, Tan gave new insight into this difficult problem by deriving
a set of relations which link the short-distance, large momentum correlations
to the bulk thermodynamic properties of a fermion system with short-range
interparticle interactions \cite{tan1,tan2,tan3,braaten08,zhang,combescot,taylor,son,braaten10,hldepl10,wernerpreprint,schakel,braatenreview,iskin,dellostritto,braaten2011}.
These relations generically apply to any dilute Fermi gas with an
inter-particle spacing much larger than $r_{0}$. All of these Tan
relations are exact in the limit of a vanishingly small range of the
interaction potential, so that $k_{F}r_{0}\rightarrow0$, where $k_{F}$
is the Fermi momentum. They are connected by a single coefficient
${\cal I}$, referred to as the integrated contact intensity or {}``contact''.
For instance, the pair correlation function is predicted to diverge
as ${\cal I}/(16\pi r^{2})$ at short distance $r(\gg r_{0})\rightarrow0$
\cite{tan1} and the momentum distribution will fall off as ${\cal I}/k^{4}$
at large momentum $k(\ll1/r_{0})$ \cite{tan2}. The original definition
of the contact given by Tan is, \begin{equation}
{\cal I}=\lim_{k\rightarrow\infty}k^{4}\rho_{\sigma}(k),\label{nktail}\end{equation}
where $\rho_{\sigma}(k)$ is the momentum distribution in one spin
component. The contact ${\cal I}$ is an extensive quantity and has
the unit of $Nk_{F}$, where $N$ is the total number of atoms of
the system. For a homogeneous Fermi gas, it is also convenient to
use a contact intensity, $\mathcal{C}=\mathcal{I}/V$, where $V$
is the volume of the system.

The immediate significance of the Tan relations can be seen most clearly
from Tan's adiabatic sweep theorem \cite{tan2}, \begin{equation}
\left[\frac{\partial E}{\partial\left(a_{s}^{-1}\right)}\right]_{S,N}=-\frac{\hbar^{2}{\cal I}}{4\pi m},\end{equation}
 which states that the rate of adiabatic change of the total energy
with respect to the inverse scattering length is proportional to the
contact. While it is surprising that the short-range, high-energy
correlations can be related so seamlessly to the low-energy equation
of state, the Tan relations are closely related to the Feynman-Hellman
theorem\cite{Hellmann1933-Feynman1939}. This method has also been
used to calculate correlations in one-dimensional Bose gases at finite
temperature from thermodynamic properties\cite{1D_BEC_Correlations},
with experimental verification\cite{Tolra-NIST-exp}.

In the fermionic strongly interacting regime, universal thermodynamics
implies a universal contact. The determination of the contact therefore
provides a very important means of characterizing the many-body properties
and phases of strongly interacting fermions, complementary to existing
measurements of the equation of state. As it is related to the derivative
of the observed energy, the contact is a more sensitive test of theoretical
predictions than direct thermodynamic measurements. We note that some
of the Tan relations have been confirmed experimentally at JILA in
the USA \cite{jilaTan} and at Swinburne University of Technology
in Australia \cite{swinTan}.

\begin{figure}[htp]
\begin{centering}
\includegraphics[clip,width=0.45\textwidth]{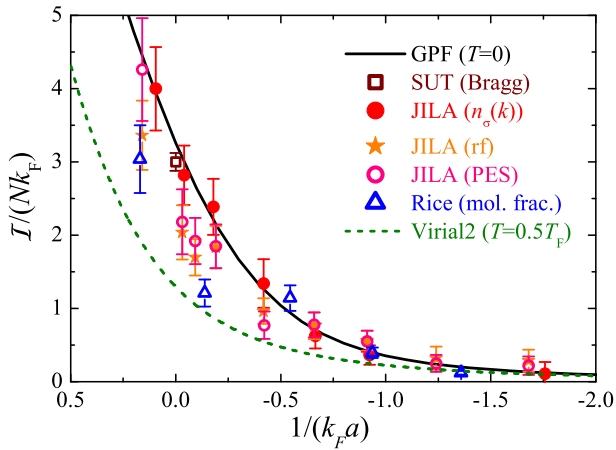} 
\par\end{centering}

\caption{(Color online) The experimental measured contact for a trapped interacting
Fermi gas (symbols) is shown as a function of the dimensionless coupling
constant $1/(k_{F}a_{s})$. It is compared with a zero-temperature
perturbation prediction (solid line) \cite{swinTan} and a finite-temperature
second-order virial expansion calculation at $0.5T_{F}$ (dashed line)
\cite{hldepl10}. The measurements include Bragg spectroscopy (empty
squares) \cite{swinTan}, momentum distributions (solid or empty circles)
\cite{jilaTan}, rf spectroscopy (stars) \cite{jilaTan}, and molecular
fraction (triangles) \cite{wernerEPJB}.}

\label{fig1} 
\end{figure}

Experimentally, the contact is obtained by a number of methods, including
measuring:
\begin{itemize}
\item the molecular fraction \cite{riceMolFrac,wernerEPJB}, 
\item the momentum distribution tail \cite{jilaTan}, 
\item the radio-frequency (rf) spectroscopy signal at large frequencies
\cite{jilaTan,punk}, or 
\item the spin-antiparallel static structure factor at large momenta, using
two-photon Bragg spectroscopy \cite{hldepl10,swinTan}. 
\end{itemize}
Figure 1 summarizes all the measured contact data to date for a trapped
interacting Fermi gas (symbols). The experimental results are compared
with our earlier predictions of a zero-temperature result from an
approximate Gaussian pair fluctuation theory (solid line) \cite{swinTan}
and from a finite-temperature calculation using the quantum virial
expansion (dashed line) \cite{hldepl10}. The experimental data of
course are \emph{not }taken exactly at zero temperature. Accurate
determination of temperature is difficult due to the absence of an
efficient thermometry. A typical experimental estimate gives $T/T_{F}\simeq0.10-0.20$.
The data in Fig. 1 agree reasonably well with the zero-temperature
curve, although near the unitarity limit they seem to lie systematically
lower, possibly due to the nonzero temperature in these experiments.
A similar theoretical calculation at zero temperature has been reported
by Werner and collaborators \cite{wernerEPJB} by interpolating the
known perturbation results in the BEC and BCS limits.

In this paper, following the pioneering studies by Yu, Palestini and
co-workers, we systematically investigate the {\em temperature}
dependence of the contact for the most interesting case of a unitary
Fermi gas. Yu \textit{et al.} estimated the finite-temperature behavior
of the universal contact \cite{yu}, based on an assumption of phonon
excitations at low temperatures and a second-order quantum virial
expansion at high temperatures; while Palestini \textit{et al.} presented
ab-initio results using a non-self-consistent \textit{T}-matrix approximation
\cite{palestini}. Both studies predicted a pronounced maximum in
the contact at finite temperatures of about $0.6T_{F}$ and $0.2T_{F}$,
respectively. The present investigation, however, calls into question
this prediction of a temperature-dependent peak in the universal contact.
Our main results are summarized below, following a two-fold motivation.

Firstly, we give a high-temperature quantum virial expansion study
of the contact up to third order. This provides us a general idea
of how the virial expansion converges with decreasing temperature,
and therefore enables us to estimate the temperature window for its
applicability. We find that the virial expansion of the contact is
{\em quantitatively} accurate down to the Fermi degenerate temperature
$T_{F}$ and $0.5T_{F}$ for a homogeneous and trapped Fermi gas at
unitarity, respectively. We show that, in exact analogy with the virial
expansion of thermodynamic potential \cite{lhdve09}, the contact
at unitarity can also be expanded in terms of universal, temperature
independent coefficients, $c_{n}=\partial\Delta b_{n}/\partial(\lambda/a_{s})$,
referred to as the contact virial coefficients. Here, $\Delta b_{n}$
is the $n$-th virial coefficient and $\lambda$ is the thermal wavelength.
This is a direct consequence of fermionic universality. We predict
that for a uniform unitary Fermi gas, the second and third contact
coefficients are given by, respectively, $c_{2,\infty}=1/\pi\simeq0.318$
and $c_{3,\infty}\simeq-0.141$. The subscript {}``$\infty$'' stands
for the unitarity limit with $a_{s}=\pm\infty$.

Secondly, we are interested in the low-temperature behavior of the
universal contact. In the strongly interacting regime, as there is
no controllable small interaction parameter, a {\em comparative}
study using different strong coupling theories is necessary \cite{hldcmp06,hldcmp08,hldcmp10}.
We find that the theoretical descriptions of the finite-temperature
contact from different strong coupling theories show considerable
discrepancies near the critical temperature $T_{c}$ ($\sim0.2T_{F}$)
for the onset of superfluidity. In particular, the enhancement (peak
structure) of the contact near $T_{c}$, found earlier in a non-self-consistent
\textit{T}-matrix calculation \cite{palestini}, is completely absent
in other strong coupling theories. \emph{Therefore, we conjecture
that the universal contact decreases monotonically with increasing
temperature, unless the temperature is much smaller than the critical
temperature.} This discrepancy remains to be resolved by future experiments
and advanced quantum Monte Carlo simulations \cite{bulgac,burovski,akkineni},
exploring the critical temperature regime close to resonance.

Our paper is organized as follows: in Sec. II we present the quantum
virial expansion for the contact and calculate the universal contact
coefficient at unitarity, for both trapped and homogeneous Fermi gases.
We then discuss how to determine the contact from strong-coupling
theories in Sec. III, using different means via the tail of momentum
distribution, Tan's adiabatic sweep relation and pressure relation.
Theoretical predictions for the uniform universal contact from different
strong-coupling theories are compared with each other and with accurate
virial expansion results in the high-temperature regime. In the next
section (Sec. IV), we describe how to calculate the trapped universal
contact from the homogeneous data and report the results of different
strong-coupling theories and virial expansion. In Sec. V, we discuss
briefly the measurement of the temperature dependence of the universal
contact in a homogeneous Fermi gas. A summary and outlook is given
in Sec. VI.

\section{Quantum virial expansion of the contact}

The quantum virial expansion has proved to be an efficient method
for studying the high-temperature properties of ultracold atomic Fermi
gases \cite{lhdve09,hmve04,dsfve10,akwve10,lhdpra10,lhdprb10,lhpreprint10}.
This method utilizes the fact that in the high temperature limit,
the chemical potential $\mu$ diverges to $-\infty$ and the fugacity
$z\equiv\exp(\mu/k_{B}T)\ll1$ is a well-defined small expansion parameter.
Thus, one may expand any physical quantities of interest in powers
of fugacity, no matter how strong the interaction is. 

For the grand thermodynamic potential of a two-component Fermi gas,
the expansion is given by \cite{lhdve09,lhdpra10}, \begin{equation}
\Omega=-k_{B}TQ_{1}\left[z+b_{2}z^{2}+\cdots+b_{n}z^{n}+\cdots\right],\end{equation}
 where $Q_{n}=Tr_{n}[\exp(-{\cal H}/k_{B}T)]$ is the partition function
of a cluster containing $n$ particles and $b_{n}$ is the $n$-th
virial expansion coefficient. The trace $Tr_{n}$ takes into account
all the quantum states of $n$-particles with a proper symmetry. The
virial coefficient $b_{n}$ can be calculated from the cluster partition
functions \cite{lhdve09,lhdpra10}, for instance, \begin{equation}
b_{2}=Q_{2}/Q_{1}-Q_{1}/2\end{equation}
 and \begin{equation}
b_{3}=Q_{3}/Q_{1}-Q_{2}+Q_{1}^{2}/3,\quad etc.\end{equation}
In practice, it is often convenient to separate out a background,
non-interacting thermodynamic potential, $\Omega^{(1)}=-k_{B}TQ_{1}[z+b_{2}^{(1)}z^{2}+\cdots+b_{n}^{(1)}z^{n}+\cdots]$.
Here, we have used the superscript {}``$1$'' to indicate the non-interacting
systems. The thermodynamic potential of interacting Fermi gases can
then be rewritten as, \begin{equation}
\Omega=\Omega^{(1)}-k_{B}TQ_{1}\left[\Delta b_{2}z^{2}+\cdots+\Delta b_{n}z^{n}+\cdots\right],\label{OmegaVE}\end{equation}
 where $\Delta b_{n}=b_{n}-b_{n}^{(1)}$. For a homogeneous Fermi
gas, $Q_{1}=2V/\lambda^{3}$ with volume $V$ and thermal wavelength
$\lambda\equiv[2\pi\hbar^{2}/(mk_{B}T)]^{1/2}$ . 

The essential assumption of the quantum virial expansion is that the
above expansion of $\Omega-\Omega^{(1)}$ converges for temperatures
down to $T_{c}$, the critical temperature for the onset of superfluidity.
We note that the fugacity in the expansion may be larger than unity
at these low temperatures. However, convergence is still possible
given small enough virial coefficients $\Delta b_{n}$. For a homogeneous
unitary Fermi gas, the second virial coefficient $\Delta b_{2,\infty}=1/\sqrt{2}$
was known 70 years ago \cite{beth1937}. The third virial coefficient,
$\Delta b_{3,\infty}\simeq-0.35501298$, was recently calculated by
the present authors \cite{lhdve09} and confirmed experimentally in
an accurate thermodynamic measurement by Nascimbène  and co-workers
\cite{ensEoS1}.

The virial expansion of the contact follows directly from an alternative
representation of Tan's adiabatic sweep theorem in the {\em grand-canonical}
ensemble, \begin{equation}
\left[\frac{\partial\Omega}{\partial\left(a_{s}^{-1}\right)}\right]_{T,\mu}=-\frac{\hbar^{2}{\cal I}}{4\pi m}.\label{adsweep2}\end{equation}
This is simply because the adiabatic sweep theorem implies the first
law of thermodynamics, \begin{equation}
\Delta E=\hbar^{2}{\cal I}/(4\pi m)\Delta(-a_{s}^{-1})+T\Delta S+\mu\Delta N\,,\end{equation}
which can alternatively be written as \begin{equation}
\Delta\Omega=\hbar^{2}{\cal I}/(4\pi m)\Delta(-a_{s}^{-1})-S\Delta T-N\Delta\mu\,.\end{equation}
Therefore, using Eq. (\ref{OmegaVE}) we immediately obtain a quantum
virial expansion for the contact: \begin{equation}
{\cal I}=\frac{4\pi m}{\hbar^{2}}k_{B}TQ_{1}\lambda\left[c_{2}z^{2}+\cdots+c_{n}z^{n}+\cdots\right],\end{equation}
 where we have defined the dimensionless contact coefficient, $c_{n}\equiv\partial\Delta b_{n}/\partial(\lambda/a_{s})$.
For a homogeneous system, we sometime use the contact intensity, $\mathcal{C}=\mathcal{I}/V$. 

In general, the contact coefficient should be a function of $\lambda/a_{s}$
and hence is temperature dependent. In the unitarity limit where $\lambda/a_{s}=0$,
however, we anticipate a constant, universal contact coefficient,
similar to the universal virial coefficient $\Delta b_{n,\infty}$
\cite{lhdve09,hmve04}. This is a manifestation of fermionic universality,
shared by all systems of strongly interacting fermions \cite{houniversality,hdlnatphys}.

\subsection{Universal relation between the homogeneous and trapped contact coefficients}

In exact analogy with the virial coefficient \cite{lhdve09}, fermionic
universality leads to a very simple relation between the trapped and
homogeneous contact coefficients at unitarity. Let us consider the
contact of a harmonically trapped Fermi gas with the trapping potential,
$V_{T}({\bf r})=m\omega^{2}r^{2}/2$ or $V_{T}({\bf r})=m\omega_{\perp}^{2}(x^{2}+y^{2})/2+m\omega_{z}^{2}z^{2}/2$
with $\omega\equiv(\omega_{\perp}^{2}\omega_{z})^{1/3}$. In the thermodynamic
limit of $\omega\rightarrow0$, we may use the local density approximation
and neglect the discrete energy levels. The whole Fermi system is
treated as many cells with a local chemical potential $\mu({\bf r})=\mu-V_{T}({\bf r})$
and a local fugacity $z(r)=e^{\mu({\bf r})/k_{B}T}\equiv z\exp(-V_{T}/k_{B}T)$.
Due to the constant contact coefficients, the spatial integration
in the total contact ${\cal I}_{T}=\int d{\bf r}[\mathcal{C}({\bf r})]$
can be easily performed. Here, we use the subscript {}``$T$'' to
indicate explicitly the trapped system. We find that, \begin{equation}
{\cal I}_{T}=\frac{4\pi m}{\hbar^{2}}k_{B}TQ_{1,T}\lambda\left[c_{2,\infty,T}z^{2}+c_{3,\infty,T}z^{3}+\cdots\right],\end{equation}
 where the trapped contact coefficient is given by a universal relation,
\begin{equation}
c_{n,\infty,T}=\frac{c_{n,\infty}}{n^{3/2}},\label{UniversalRelation}\end{equation}
 and $Q_{1,T}=2(k_{B}T)^{3}/(\hbar\omega)^{3}$ is the single-particle
partition function in harmonic traps and in the local density approximation.

In the following, using the known solution of two- and three-fermion
problems, we calculate the universal second and third contact coefficients,
in both homogeneous and trapped configurations.

\subsection{Second universal contact coefficient}

The second contact coefficient of a {\em homogeneous} interacting
Fermi gas can be obtained from the well-known phase-shift expression
for the second virial coefficient \cite{hmve04,beth1937}, \begin{equation}
\frac{\Delta b_{2}}{\sqrt{2}}=\sum_{i}e^{-E_{b}^{i}/(k_{B}T)}+\frac{1}{\pi}\int\limits _{0}^{\infty}dk\frac{d\delta_{0}}{dk}e^{-\lambda^{2}k^{2}/\left(2\pi\right)}.\end{equation}
 Here, $E_{b}^{i}$ is the energy of the $i$-th bound state of the
two-body attractive interaction and $\delta_{0}(k)$ is the \textit{s}-wave
scattering phase shift, given by $k\cot\delta_{0}\simeq-1/a_{s}+r_{0}k^{2}/2$.
To obtain the derivative $\partial\Delta b_{2}/\partial(\lambda/a_{s})$,
let us choose the BCS side with $a_{s}<0$, on which the two-body
bound state is absent. As the range of the interaction potential $r_{0}$
is vanishingly small so that $k_{B}T\ll\hbar^{2}/(mr_{0}^{2})$ and
$k\ll1/r_{0}$, we find that $d\delta_{0}(k)/dk\simeq-a_{s}/[1+k^{2}a_{s}^{2}]$
and, \begin{equation}
\Delta b_{2}(a_{s}<0)=\frac{\sqrt{2}}{\pi}\int\limits _{0}^{\infty}dt\frac{1}{1+t^{2}}e^{-\lambda^{2}t^{2}/\left(2\pi a_{s}^{2}\right)}.\end{equation}
 Near to the resonance, it is readily seen that $\Delta b_{2}\simeq1/\sqrt{2}+\lambda/(\pi a_{s})$,
giving rise to a homogeneous contact coefficient, \begin{equation}
c_{2,\infty}=\frac{1}{\pi}.\end{equation}

To calculate the {\em trapped} second contact coefficient, we consider
the second virial sufficient in an isotropic harmonic trap, which
is given by \cite{lhdve09,lhdpra10}, \begin{equation}
\Delta b_{2,T}=\frac{1}{2}\sum_{n}\left[e^{-\epsilon_{rel,n}/k_{B}T}-e^{-\epsilon_{rel,n}^{(1)}/k_{B}T}\right].\end{equation}
 Here, $\epsilon_{rel,n}=(2\nu_{n}+3/2)\hbar\omega$ is the $n$-th
relative energy of two fermions with unlike spins and $\nu_{n}$ satisfies
the secular equation, $2\Gamma(-\nu_{n})/\Gamma(-\nu_{n}-1/2)=d/a_{s}$,
with $\Gamma$ the gamma function and $d=\sqrt{2\hbar/(m\omega)}$
the characteristic length scale of the harmonic trap. The non-interacting
relative energy is $\epsilon_{rel,n}^{(1)}=(2n+3/2)\hbar\omega$ ($n=0,1,2,...$)
and in the unitarity limit, the solution of $\nu_{n}$ is known analytically,
$\nu_{n,\infty}=n-1/2$. It is easy to show that, \begin{equation}
\left[\frac{\partial\epsilon_{rel,n}}{\partial\left(\lambda/a_{s}\right)}\right]_{\lambda/a_{s}=0}=-\frac{\hbar\omega d}{\pi\lambda}\frac{\Gamma\left(n+1/2\right)}{n!}.\end{equation}
 Thus, we find that in the unitarity limit, \begin{equation}
c_{2,\infty,T}=\frac{d}{2\pi\lambda}\tilde{\omega}\sum_{n=0}^{\infty}\frac{\Gamma\left(n+1/2\right)}{n!}e^{-\left(2n+1/2\right)\tilde{\omega}},\end{equation}
 where $\tilde{\omega}\equiv\hbar\omega/(k_{B}T)\ll1$ is the reduced
trapping frequency. The sum over $n$ can be exactly performed, leading
to, \begin{eqnarray}
c_{2,\infty,T} & = & \frac{1}{2\sqrt{2}\pi}\left[\frac{2\tilde{\omega}}{e^{+\tilde{\omega}}-e^{-\tilde{\omega}}}\right]^{1/2}\nonumber \\
 & = & \frac{1}{2\sqrt{2}\pi}\left[1-\frac{\tilde{\omega}^{2}}{12}+O\left(\tilde{\omega}^{4}\right)\right].\end{eqnarray}
 The leading term in the above equation is universal, satisfying the
universal relation Eq. (\ref{UniversalRelation}). The second term
($\propto\tilde{\omega}^{2}$) is non-universal and is caused by the
length scale of the harmonic trap \cite{lhdve09}. It represents the
finite-size correction to the local density approximation that we
have adopted above. This correction however is extremely small if
$\tilde{\omega}\rightarrow0$, as anticipated. At the Fermi degenerate
temperature, $T_{F}=(3N)^{1/3}\hbar\omega/k_{B}$, we find that $\tilde{\omega}^{2}\propto N^{-2/3}\sim10^{-4}$,
for the typical number of atoms $N\sim10^{5}$ in the current experiments
\cite{rmpgiorgini}.

\subsection{Third universal contact coefficient}

The determination of the third contact coefficient is more cumbersome.
In the homogeneous case, it is extremely difficult to use the conventional
method employing the three-particle scattering matrix to obtain the
third virial coefficient. Therefore, it is more practical to determine
firstly the trapped contact coefficient $c_{3,\infty,T}$, and then
to use the universal relations at low trap frequency to obtain the
homogeneous result, $c_{3,\infty}=3\sqrt{3}c_{3,\infty,T}$. We note
that this method relies on exact three-body solutions to the energy
eigenvalues in a trap. These are known analytically for all eigenstates,
and can be summed numerically to any required accuracy.

An estimate of $c_{3,\infty,T}$ can already be obtained by the known
results of $\Delta b_{3,T}$ as a function of the coupling constant
$1/k_{F}a_{s}$ at different temperatures $T/T_{F}$ and at a small
$\tilde{\omega}\sim0.1$ (see, for example, Fig. 3\textit{b} in Ref.
\cite{lhdve09}), simply because, \begin{equation}
c_{3,T}\equiv\frac{1}{k_{F}\lambda}\frac{\partial\Delta b_{3,T}}{\partial(1/k_{F}a_{s})}=\sqrt{\frac{T}{4\pi T_{F}}}\frac{\partial\Delta b_{3,T}}{\partial(1/k_{F}a_{s})}.\end{equation}
 We find that the coefficient $c_{3,\infty,T}$ at resonance is indeed
nearly temperature independent and estimate from the slope of $\Delta b_{3,T}$
that, $c_{3,\infty,T}($estimate$)\simeq-0.0265$ at $\tilde{\omega}\sim0.1$.

An accurate determination of $c_{3,\infty,T}$ requires a systematic
extrapolation to the limit of $\tilde{\omega}=0$. For this purpose,
we calculate numerically the derivative $c_{3,\infty,T}(\tilde{\omega})=[\partial\Delta b_{3,T}/\partial(\lambda/a_{s})]_{\lambda/a_{s}=0}\ $as
a function of $\tilde{\omega}$. Using the small $\tilde{\omega}$
data, a numerical extrapolation to $\tilde{\omega}=0$ gives rise
to the trapped third virial contact coefficient, \begin{equation}
c_{3,\infty,T}\simeq-0.0271.\end{equation}
Thus, we obtain immediately from the universal relation, Eq. (\ref{UniversalRelation}),
the homogeneous third contact coefficient, \begin{equation}
c_{3,\infty}=-0.141.\end{equation}

\subsection{Large-$T$ contact: the homogeneous case}

We are now ready to calculate the universal contact in the high temperature
regime. For a homogeneous Fermi system, the single-particle partition
function $Q_{1}=2V/\lambda^{3}$ and the dimensionless contact ${\cal I}/(Nk_{F})$
is given by, \begin{equation}
\frac{{\cal I}}{Nk_{F}}\equiv\frac{\mathcal{C}}{\rho k_{F}}=3\pi^{2}\left(\frac{T}{T_{F}}\right)^{2}\left[c_{2,\infty}z^{2}+c_{3,\infty}z^{3}+\cdots\right].\label{homContactVE}\end{equation}
Here $N\equiv\rho V$ is the total number of atoms with the homogeneous
density $\rho$.

The fugacity $z$ is determined by the number equation \cite{lhdpra10},
\begin{equation}
\tilde{\rho}=\tilde{\rho}^{(1)}\left(z\right)+\left[2\Delta b_{2,\infty}z^{2}+3\Delta b_{3,\infty}z^{3}+\cdots\right],\label{homNumVE}\end{equation}
 where we have defined a dimensionless density $\tilde{\rho}\equiv\rho\lambda^{3}/2=[4/(3\sqrt{\pi})](T_{F}/T)^{3/2}$
and the density of a non-interacting Fermi gas as \begin{equation}
\tilde{\rho}^{(1)}(z)\equiv(2/\sqrt{\pi})\int_{0}^{\infty}dt\sqrt{t}/(1+z^{-1}e^{t})\,.\end{equation}

In practice, for a given fugacity, we calculate the dimensionless
density using Eq. (\ref{homNumVE}) and hence the reduced temperature
$T/T_{F}$. The dimensionless contact is then obtained from Eq. (\ref{homContactVE}),
as a function of $T/T_{F}$ or the inverse fugacity $z^{-1}$.

\begin{figure}[htp]
\begin{centering}
\includegraphics[clip,width=0.45\textwidth]{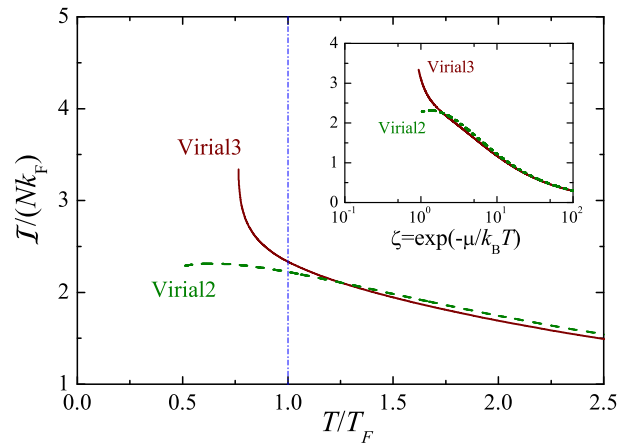} 
\par\end{centering}

\caption{(Color online) Universal contact of a homogeneous unitary Fermi gas
at high temperatures, as predicted by the quantum virial expansion
method up to the second order (dashed line) and the third order (solid
line). Dashed vertical line indicates the Fermi degenerate temperature
$T_{F}$. The inset shows the contact as a function of the inverse
fugacity.}

\label{fig2} 
\end{figure}

Figure 2 reports the temperature (main figure) or fugacity (inset)
dependence of the homogeneous contact in the unitarity limit, calculated
by virial expanding to the second order (dashed line) or third order
(solid line). The close agreement between the second and third predictions
strongly indicates that the virial expansion works {\em quantitatively}
well down to the Fermi degenerate temperature $T_{F}$, as indicated
by the vertical dashed line. At sufficient high temperatures, where
\begin{equation}
z\simeq\tilde{\rho}=[4/(3\sqrt{\pi})](T_{F}/T)^{3/2}\,,\label{eq:homogeneous density}\end{equation}
the leading temperature dependence of the contact is given by, \begin{equation}
\frac{{\cal I}}{Nk_{F}}\left(T\gg T_{F}\right)=\frac{16}{3}\left(\frac{T}{T_{F}}\right)^{-1},\end{equation}
 as predicted by Yu and co-workers \cite{yu}. We note however that
the prefactor there is smaller by a factor of $4\pi^{2}$, due to
a different definition for the contact.

\subsection{Large-$T$ contact: the trapped case}

For a trapped Fermi gas at unitarity, the dimensionless contact can
be written as, \begin{equation}
\frac{{\cal I}_{T}}{Nk_{F}}=24\pi^{^{3/2}}\left(\frac{T}{T_{F}}\right)^{7/2}\left[c_{2,\infty,T}z^{2}+c_{3,\infty,T}z^{3}+\cdots\right].\label{trapContactVE}\end{equation}
 The number equation takes the form \cite{lhdpra10}, \begin{equation}
\tilde{\rho}_{T}=\tilde{\rho}_{T}^{(1)}\left(z\right)+\left[2\Delta b_{2,\infty,T}z^{2}+3\Delta b_{3,\infty,T}z^{3}+\cdots\right],\label{trapNumVE}\end{equation}
 where $\tilde{\rho}_{T}\equiv(N/2)(\hbar\omega)^{3}/(k_{B}T)^{3}=(T_{F}/T)^{3}/6$
and the density of a non-interacting trapped Fermi gas $\tilde{\rho}_{T}^{(1)}(z)\equiv(1/2)\int_{0}^{\infty}dtt^{2}/(1+z^{-1}e^{t})$.
The trapped virial coefficient is given by $\Delta b_{n,\infty,T}=\Delta b_{n,\infty}/n^{3/2}$.
In analogy with the homogeneous case, for a given fugacity we determine
the reduced temperature $T/T_{F}$ from the number equation (\ref{trapNumVE})
and then calculate the trapped contact using Eq. (\ref{trapContactVE}).

\begin{figure}[htp]

\begin{centering}
\includegraphics[clip,width=0.45\textwidth]{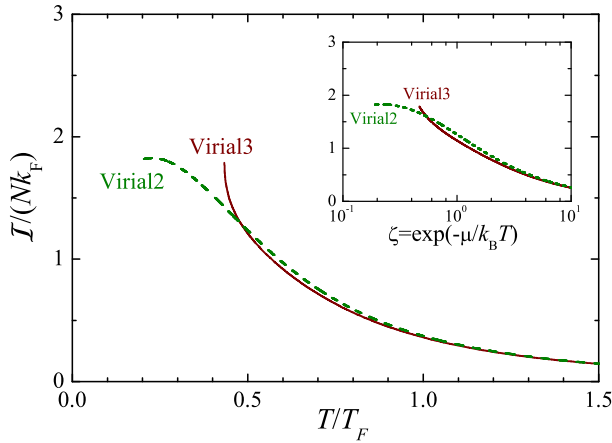} 
\par\end{centering}

\caption{(Color online) Universal contact of a trapped unitary Fermi gas at
high temperatures, obtained by expanding the virial series to the
second order (dashed line) and the third order (solid line). The inset
shows the contact as a function of the inverse fugacity.}

\label{fig3} 
\end{figure}

Figure 3 presents the virial expansion prediction for the trapped
universal contact, expanding up to the second order (dashed line)
or third order (solid line). Amazingly, because of the factor of $n^{-3/2}$
reduction for the $n$-th contact coefficient in harmonic traps, the
convergence of the expansion is much improved. The expansion now seems
to be quantitatively reliable down to $0.5T_{F}$. The asymptotic
behavior of the contact at very high temperatures can be determined
by setting \begin{equation}
z\simeq\tilde{\rho}_{T}=(T_{F}/T)^{3}/6\,.\label{eq:trapped density}\end{equation}
We find that, \begin{equation}
\left(\frac{{\cal I}}{Nk_{F}}\right)_{T}\left(T\gg T_{F}\right)=\frac{\sqrt{2\pi}}{6}\left(\frac{T}{T_{F}}\right)^{-5/2},\end{equation}
 in agreement with our previous result from the large momentum behavior
of static structure factor \cite{hldepl10}. Thus, the contact in
harmonic traps decays at high temperatures much faster than in homogeneous
space, due to the reduction of the peak density at the trap center
at high temperatures.

\section{Low-$T$ homogeneous contact from strong coupling theories}

At low temperatures, we have to resort to strong coupling theories
to determine the contact. In the absence of a controllable small interaction
parameter, however, there is no known {\em a priori} theoretical
justification for which strong coupling theory is the most appropriate
\cite{hldcmp10}. Here we consider three different theories within
the many-body \textit{T}-matrix approximation: the Nozières and Schmitt-Rink
(NSR) Gaussian pair fluctuation theory \cite{nsr,sademelo,hldepl06,lhepl06,diener},
non-self-consistent $G_{0}G_{0}$ theory \cite{strinati02,strinati04,combescotPRA},
and self-consistent $GG$ theory \cite{gg94,lhgg05,gg07}.

\subsection{Brief review of $T$-matrix approximations}

The $T$-matrix approximation is also referred to as ladder approximation,
which involves infinite set of Feynman diagrams --- the ladder sum
in the particle-particle channel. It is now generally accepted that
this ladder sum is necessary for accounting for strong pair fluctuations
in the strongly interacting regime and is the leading class of all
sets of diagrams \cite{prloketev,pryanase}. The need for including
infinite sets of Feynman diagrams is self-evident, since we are dealing
with a strongly interacting theory in which there is no small coupling
constant that would allow a truncation of perturbation theory to finite
order.

There are several $T$-matrix theories, differing in the diagrammatic
structure of the \textit{T}-matrix $t(Q)$, the particle-particle
propagator $\chi\left(Q\right)$, and the self-energy $\Sigma\left(K\right)$.
The ones we consider here are respectively given by the following
form \cite{prloketev,pryanase}, \begin{eqnarray}
t(Q) & = & \frac{U}{\left[1+U\chi\left(Q\right)\right]},\label{tmatrix}\\
\chi\left(Q\right) & = & \sum\nolimits _{K}G_{\alpha}\left(K\right)G_{\alpha}\left(Q-K\right),\label{pp2p}\\
\Sigma\left(K\right) & = & \sum\nolimits _{Q}t\left(Q\right)G_{\alpha}\left(Q-K\right).\label{selfenergy}\end{eqnarray}
 Here and throughout, $Q=({\bf q},i\nu_{n})$, $K=({\bf k},i\omega_{m})$,
while $U^{-1}=m/(4\pi\hbar^{2}a_{s})-\sum_{{\bf k}}1/(2\epsilon_{{\bf k}})$
is the {\em bare} contact interaction renormalized in terms of
the \textit{s}-wave scattering length. We follow the conventional
notations and use $\epsilon_{{\bf k}}=\hbar^{2}{\bf k}^{2}/(2m)$
and $\sum_{K}=$ $k_{B}T\sum_{m}\sum_{{\bf k}}$, where ${\bf q}$
and ${\bf k}$ are wave vectors, while $\nu_{n}$ and $\omega_{m}$
are bosonic and fermionic Matsubara frequencies. The Greek subscript
$\alpha$ in the expressions for the propagator $\chi\left(Q\right)$
and the self-energy $\Sigma\left(K\right)$ may either be set to {}``0'',
indicating a non-interacting Green's function $G_{0}(K)$, or be absent,
indicating a full interacting Green's function $G\left(K\right)$.
The latter can be calculated once the self-energy is determined, using
the Dyson equation, \begin{equation}
G\left(K\right)=\frac{G_{0}\left(K\right)}{\left[1-G_{0}\left(K\right)\Sigma\left(K\right)\right]}.\label{dysoneq}\end{equation}
 By taking different values of $\alpha$, two different $T$-matrix
approximations are obtained: either self-consistent (no $\alpha$),
or non-selfconsistent ($\alpha=0$) . We note that, in the superfluid
phase the Green's function has to be a $2\times2$ matrix, accounting
for the $U(1)$ symmetry breaking.

The simplest choice in the $T$-matrix approximation is to take a
non-interacting Green function $G_{0}(K)$ everywhere in $\chi\left(Q\right)$
and $\Sigma\left(K\right)$. This idea was pioneered by Nozières and
Schmitt-Rink (NSR) for an interacting Fermi gas in its normal state
\cite{nsr}. In addition, an approximate Dyson equation for the Green's
function, \textit{i.e.}, \begin{equation}
G\left(K\right)=G_{0}\left(K\right)+G_{0}\left(K\right)\Sigma\left(K\right)G_{0}\left(K\right),\label{gfnsr}\end{equation}
 was used. Interestingly, this non-self-consistent NSR approach combined
with a truncated Dyson equation can be reformulated using the functional
path-integral language for the grand thermodynamic potential \cite{sademelo},
expanded to second order to include Gaussian fluctuations. From this
perspective, the NSR theory is the first step in an order-by-order
expansion in path-integral fluctuations, which has a systematic path
to including higher-order terms.

The NSR theory was extended recently to the broken-symmetry superfluid
phase by several authors in different manners \cite{hldepl06,diener,ohashiprl02,ohashipra03,footnote},
some of which involve assumptions in order to ease the computational
workload. A full extension following the original NSR approach was
reported by the present authors in 2006 \cite{hldepl06}, with the
use of a mean-field ($2\times2$ matrix) BCS Green's function as {}``$G_{0}$''.
Hereafter we shall refer to this extension as a Gaussian pair fluctuation
theory or the GPF approach.

The truncation of the Dyson equation, Eq. (\ref{gfnsr}), can be avoided,
as shown by Strinati \cite{strinati02,strinati04} and Combescot \cite{combescotPRA}
and their co-workers. In particular, in the superfluid state the coupled
$T$-matrix equations have been solved \cite{strinati04}. In the
following, we shall refer to this non-self-consistent $T$-matrix
approximation with no truncation in the Dyson equation as the $G_{0}G_{0}$
theory. However, since this approach is no longer part of a systematic
expansion of the path integral, it is an open question as to whether
it is more or less accurate than the NSR theory.

More sophisticated strong-coupling theories can also be obtained by
using the full (dressed) Green function $G\left(K\right)$ in $\chi\left(Q\right)$
and $\Sigma\left(K\right)$. This modification, referred to as the
self-consistent $GG$ theory, was investigated in detail at the BEC-BCS
crossover by Haussmann and collaborators \cite{gg94,gg07}, above
and below the superfluid transition temperature. One advantage of
the $GG$ approximation is that the theory satisfies the so-called
$\Phi-$derivability and is thus conserving, although the conservation
is at the single-particle level only. The $GG$ theory has also been
intensively studied in the condensed matter physics, particularly
for high-temperature materials, under the name of the fluctuation
exchange (FLEX) approximation \cite{prloketev,pryanase}. Below threshold
the GG approximation is difficult to treat at unitarity, and generally
involves additional assumptions in order to obtain a scale-invariant
theory.

It is clear that in both the NSR or $G_{0}G_{0}$ approximations one
omits infinite classes of diagrams that are responsible for multiparticle
interactions. The fully self-consistent $GG$ theory attempts to correct
for this, by modifying the single-particle Green function in the ladder
diagram. However, the more crucial interaction vertices remain unchanged.
At this stage, there are no general grounds for deciding which strong-coupling
theory is the most appropriate. Therefore, any theoretical predictions
produced by different strong coupling theories should be treated conservatively
and be examined critically using more accurate \emph{ab-initio} quantum
Monte Carlo simulations or future experiments. We note that, for the
equation of state of strongly interacting fermions, a comprehensive
comparison between different strong-coupling theories and most recent
experimental results was reported by the present authors \cite{hldcmp10}.
In this earlier comparison we also included intermediate ($GG_{0}$)
theories which are partially self-consistent\cite{ChenReview}. These
are omitted here for simplicity.

In the following, we calculate the universal contact at unitarity
by using the GPF or NSR theory and compare our results with these
reported by Strinati and co-workers ($G_{0}G_{0}$) \cite{palestini}
and with the predictions provided by Haussmann and co-workers ($GG$)
\cite{gg07,haussmann,enss}.

\subsection{Determination of the contact}

\begin{figure}[htp]

\begin{centering}
\includegraphics[clip,width=0.45\textwidth]{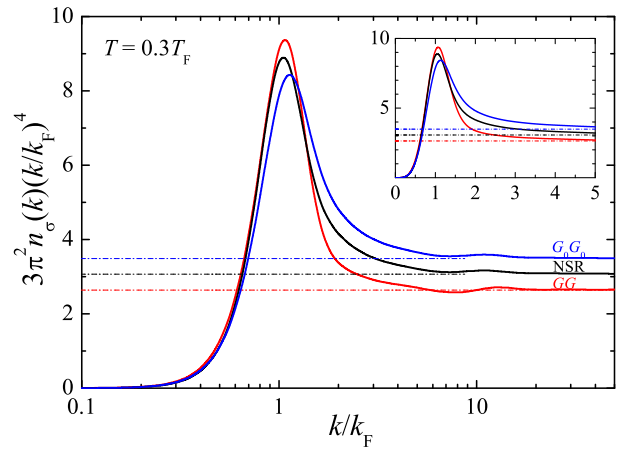} 
\par\end{centering}

\caption{(Color online) Large wave-vector power-law behavior in the momentum
distribution $\rho_{\sigma}(k)$, obtained using different strong-coupling
theories. The temperature $T=0.3T_{F}$ is larger than the critical
temperature $T_{c}\sim0.2T_{F}$. The dashed lines indicate the asymptotic
value of $3\pi^{2}k^{4}\rho_{\sigma}(k)/k_{F}^{4}$, which gives rise
to the dimensionless contact ${\cal I}/(Nk_{F})$. The inset shows
the plot in the linear scale.}

\label{fig4} 
\end{figure}

There are a number of ways to calculate the contact from $T$-matrix
theories according to the different Tan relations, as follows: 
\begin{enumerate}
\item Following Tan's original definition \cite{tan1}, the most direct
way seems to be the use of the large wave-vector asymptotic behavior
of the fermionic momentum distribution, Eq. (\ref{nktail}). The theoretical
contact from the $G_{0}G_{0}$ theory was obtained in this manner
by Strinati and co-workers \cite{palestini}. In the JILA experiment,
the contact was extracted by averaging $k^{4}\rho_{\sigma}(k)$ over
an interval at $k_{C,expt}\sim2k_{F}$ \cite{jilaTan}. Figure 4 shows
$k^{4}\rho{}_{\sigma}(k)$ for a unitary Fermi gas in its normal state
at $T=0.3T_{F}$, obtained by solving (iteratively) the coupled $T$-matrix
equations \cite{lhgg05}, Eqs. (\ref{tmatrix}), (\ref{pp2p}), (\ref{selfenergy}),
and (\ref{dysoneq}). The power-law $k^{-4}$ tail is clearly identified
since $k^{4}\rho_{\sigma}(k)$ approaches a constant above a characteristic
wave-vector $k_{C}\sim5k_{F}$, which is however larger than $k_{C,expt}$.
The small wiggles in Fig. 4 at about $10k_{F}$ are due to the numerical
inaccuracy. We note that the power-law tail of $k^{-4}$ in the momentum
distribution was first pointed out by Haussmann \cite{gg94}, using
the self-consistent $GG$ theory for a normal interacting Fermi gas.
In the $GG$ theory, the tail in $\rho_{\sigma}(k)$ implies that
the contact can be directly calculated from the order parameter $\Delta$
and the vertex function $\Gamma({\bf x},\tau)$, \begin{equation}
\frac{{\cal I}}{Nk_{F}}=\frac{3\pi^{2}}{4}\frac{\left[\Delta^{2}-\Gamma_{11}\left({\bf x}=0,\tau=0^{-}\right)\right]}{\epsilon_{F}^{2}}.\label{eq:contactVertex}\end{equation}
A brief explanation of this expression is given in the Appendix A.
We also refer to Sec. IID of Ref. \cite{haussmann} for more details.
\item Another way to calculate the contact is to use Tan's pressure relation
\cite{tan2}, \begin{equation}
PV-\frac{2}{3}E=\frac{\hbar^{2}{\cal I}}{12\pi ma_{s}},\label{pressure}\end{equation}
 which results directly from the adiabatic sweep relation. As discussed
in Ref. \cite{tan2}, this is evident if we write the total energy
in the form, $E=N\epsilon_{F}f_{E}[1/(k_{F}a_{s}),S/(Nk_{B})]$, where
$f_{E}$ is a dimensionless function. The adiabatic sweep relation
leads to, $\hbar^{2}{\cal I}/(4\pi m)=-N\epsilon_{F}f_{E}^{\prime}/k_{F}$,
where the derivative is taken with respect to the variable $1/(k_{F}a_{s})$.
We then find that, \begin{eqnarray}
PV & \equiv & -V\left[\frac{\partial E}{\partial V}\right]_{S,N,a_{s}^{-1}},\\
 & = & -N\left[V\frac{\partial\epsilon_{F}}{\partial V}\right]f_{E}-N\epsilon_{F}f_{E}^{\prime}\left[V\frac{\partial k_{F}^{-1}}{\partial V}\right]\frac{1}{a_{s}},\\
 & = & \frac{2}{3}E+\frac{\hbar^{2}{\cal I}}{12\pi ma_{s}}.\end{eqnarray}
 A direct application of the pressure relation in the unitarity limit
is prohibited as $a_{s}^{-1}=0$ and therefore the dependence on the
contact is lost. However, we can use the pressure relation to calculate
the contact around the unitarity and then do a smooth interpolation
to the limit of $1/(k_{F}a_{s})=0$.
\item Finally, the contact can also be calculated from the entropy by using,
\begin{equation}
\left(\frac{\partial{\cal I}}{\partial T}\right)=\frac{4\pi m}{\hbar^{2}}\left[\frac{\partial S}{\partial\left(-a_{s}^{-1}\right)}\right]_{T,\mu},\label{dIdT}\end{equation}
 which can be obtained directly by taking temperature derivative in
Eq. (\ref{adsweep2}). This important relation was first shown by
Yu and co-workers \cite{yu}. Thus, the temperature dependence of
the contact is determined by the variation in the entropy with respect
to the coupling constant.
\end{enumerate}
\begin{figure}[htp]
\begin{centering}
\includegraphics[clip,width=0.45\textwidth]{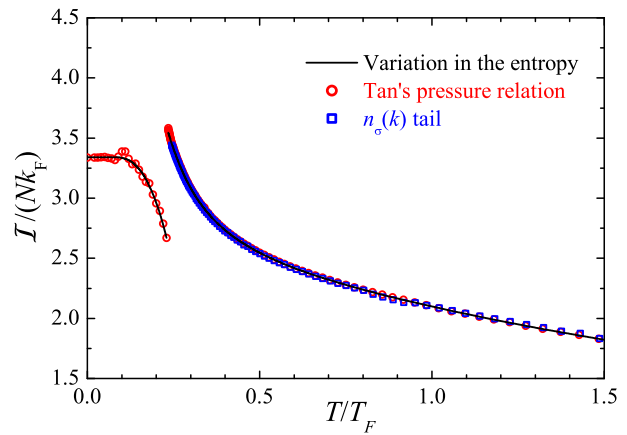} 
\par\end{centering}

\caption{(Color online) Temperature dependence of the contact in the unitarity
limit, obtained by using the variation in the entropy Eq. (\ref{dIdT}),
Tan's pressure relation Eq. (\ref{pressure}), and the momentum distribution
tail Eq. (\ref{nktail}). Here we use the Gaussian pair fluctuation
theory \cite{hldepl06}.}

\label{fig5} 
\end{figure}

Figure 5 reports the universal contact calculated using the above
mentioned methods, within the GPF (NSR) theory. There is an apparently
discontinuous behaviour at the critical temperature $T_{c}$, due
to the breakdown of the NSR approach just above $T_{c}$ \cite{hldcmp08,hldcmp10}.
We find that all three theoretical methods yield nearly the same prediction,
owing to the consistency in thermodynamic relations. The self-consistent
$GG$ theory has the same advantage of thermodynamic consistency.

\subsection{Universal homogeneous contact from strong-coupling theories}

\begin{figure}[htp]

\begin{centering}
\includegraphics[clip,width=0.45\textwidth]{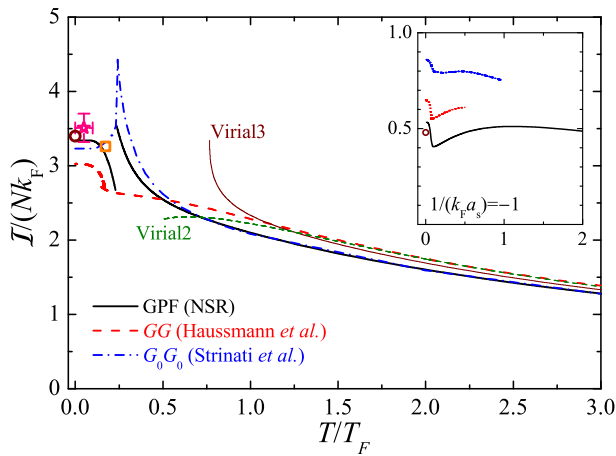} 
\par\end{centering}

\caption{(Color online) Universal contact of a homogeneous Fermi gas in the
unitarity limit. The predictions from different strong coupling theories
are compared with each other. At high temperatures, they are compared
with the virial expansion results as well. The symbols at low temperature
indicate the predictions calculated by using the measured equation
of state (star) \cite{ensEoS2}, the quantum Monte Carlo result for
pair correlation function or ground state energy (circle) \cite{combescotEPL,latestQMCT0}
and the lattice quantum Monte Carlo simulation at the critical temperature
(square). In the inset, the temperature dependence of the contact
is shown for a BCS Fermi gas at the interaction strength $1/(k_{F}a_{s})=-1$.
The circle at $T=0$ shows the contact obtained from the quantum Monte
Carlo result for pair correlation function \cite{pcfPRL}. }

\label{fig6} 
\end{figure}

\begin{table}[h]
\begin{tabular}{cc}
 & \tabularnewline
\hline
\hline 
${\cal I}/(Nk_{F})$  & Methods \tabularnewline
\hline 
$3.51\pm0.19$  & ENS equation of state data (Ref. \cite{ensEoS2}) \tabularnewline
\hline 
$3.396(9)$  & QMC ground state energy (Ref. \cite{latestQMCT0}) \tabularnewline
$3.40$  & QMC pair correlation function (Ref. \cite{combescotEPL}) \tabularnewline
$3.34$  & GPF (present work) \tabularnewline
$3.02$  & $GG$ (Ref. \cite{haussmann}) \tabularnewline
$3.23$  & $G_{0}G_{0}$ (Ref. \cite{palestini}) \tabularnewline
\hline
\end{tabular}

\caption{Zero temperature contact in the unitarity limit. The contact ${\cal I}/(Nk_{F})=6\pi\zeta/5$
can be calculated from the ground state energy of an interacting Fermi
gas near unitarity \cite{tan2}: $E/(N\epsilon_{F})=(3/5)[\xi-\zeta/(k_{F}a_{s})+\cdots]$,
where $\xi$ and $\zeta$ are two universal parameters. The contact
may also be determined from the tail of spin-antiparallel static structure
factor \cite{hldepl10}, $S_{\uparrow\downarrow}(q)={\cal I}/(4Nq)$.
Here, we compare different results obtained from the three strong-coupling
theories with the predictions by using $\zeta$, either measured at
ENS ($\zeta=0.93(5)$) \cite{ensEoS2} or extracted from quantum Monte
simulation for ground state energy ($\zeta=0.901(2)$) \cite{latestQMCT0}
or for the static structure factor (or pair correlation function ($\zeta=0.90$))
\cite{combescotEPL}.}

\end{table}

Figure 6 presents the comparison of different theoretical predictions
for the temperature dependence of the contact of a homogeneous Fermi
gas at unitarity. The zero temperature results are listed in the Table
I, as well as the contact calculated by using the low temperature
experimental data of the equation of state \cite{ensEoS2} and by
using the quantum Monte Carlo data for pair correlation function \cite{combescotEPL}
or ground state energy \cite{latestQMCT0}. The experimental data
of course is \emph{not }at zero temperature, as this is not experimentally
achievable. In these measurements, low temperature thermometry was
difficult. Similar experiments carried out elsewhere have reported
lowest achievable temperatures of around $T/T_{F}\simeq0.05$.

At low temperatures, we observe two distinct predictions for the behavior
of the contact with increasing temperature: while the $G_{0}G_{0}$
theory predicts an increase of the contact and hence a maximum around
the critical temperature $T_{c}$, both the GPF and $GG$ theory suggest
that the contact decreases monotonically as the temperature increases,
unless $T\ll T_{c}$. This qualitative discrepancy deserves a careful
analysis.

There are known arguments for the presence of a maximum in the contact
at low temperature, as follows: 
\begin{itemize}
\item The enhancement of the contact just above $T_{c}$ in the $G_{0}G_{0}$
theory was interpreted to be related to the strengthening of local
pairing correlations in the absence of long-range order \cite{palestini}. 
\item As shown by Yu and co-workers \cite{yu}, phonon excitations at low
temperatures will enhance the contact as $T^{4}$, causing a growth
in the contact with temperature. 
\item The assumption of Fermi-liquid behavior for a weakly coupled Fermi
gas \cite{yu} would also lead to a maximum in the contact at $T\sim T_{F}$. 
\end{itemize}
These arguments, however, are not convincing enough to conclude there
is a real maximum in the low-temperature contact in the unitarity
limit, as we discuss below.

The difficulty is that all the strong coupling theories are less accurate
near criticality, suffering from strong pair fluctuations. Thus, the
enhancement of the contact just above $T_{c}$ in the $G_{0}G_{0}$
theory might be related to the breakdown of the approximations in
this approach. Indeed, we observe a very similar enhancement of the
contact in the NSR theory. However, we know that the NSR approach
simply breaks down in the temperature region just above criticality.

\begin{figure}[htp]
\begin{centering}
\includegraphics[clip,width=0.45\textwidth]{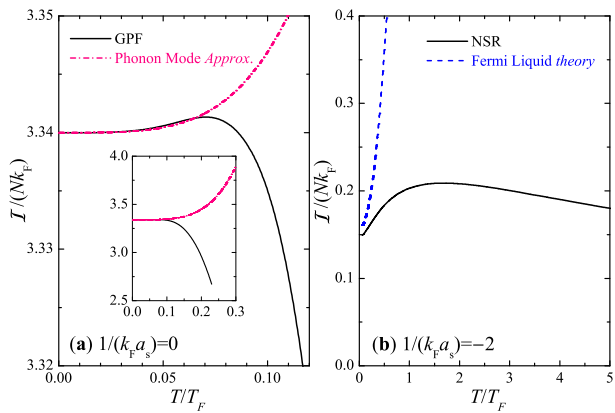} 
\par\end{centering}

\caption{(Color online) (a) In the unitarity limit, the low temperature contact
calculated from the GPF theory (solid line) is compared with the $T^{4}$
power-law of phonon excitations (dot-dashed line). The inset shows
the comparison in a larger temperature window. (b) The contact for
a weakly interacting Fermi gas at the coupling constant $1/(k_{F}a_{s})=-2$.
The NSR result (solid line) is contrasted with a prediction from the
Fermi-liquid theory (dashed line).}

\label{fig7} 
\end{figure}

On the other hand, while phonon excitations are important at low temperatures,
our realistic calculation using the GPF theory at low temperatures
suggests that the power-law of $T^{4}$ due to phonons is exhibited
at $T<0.07T_{F}\ll T_{c}$ only, as clearly shown in Fig. 7\textit{a}.
At such low temperatures, the contribution from phonons is of relative
order $10^{-3}$ . This is so nearly negligible that it might be easily
be compensated by other excitations in this strongly interacting superfluid
at unitarity.

We have also investigated the Fermi-liquid theory of the contact for
a weak-coupling Fermi gas, as reported in Fig. 7\textit{b}, where
we compare the Fermi-liquid result with the NSR prediction for a weakly
interacting Fermi gas with $k_{F}a=-0.5$. The essential idea of the
Fermi-liquid theory is that at low temperatures, the entropy density
is given by $s=m^{*}k_{F}k_{B}^{2}T/(3\hbar^{2})$, where the effective
mass $m^{*}=m[1+8(7\ln2-1)/(15\pi^{2})k_{F}^{2}a_{s}^{2}]$. Therefore,
by using Eq. (\ref{dIdT}), the derivative of the contact with temperature
is given by, \begin{equation}
\frac{\partial\left[{\cal I}/\left(Nk_{F}\right)\right]}{\partial\left(T/T_{F}\right)}=-\frac{16(7\ln2-1)\pi}{15}\left(k_{F}a_{s}\right)^{3}\frac{T}{T_{F}}>0.\end{equation}
 This leads to, \begin{equation}
\frac{{\cal I}}{Nk_{F}}=\frac{{\cal I}_{0}}{Nk_{F}}-\frac{8(7\ln2-1)\pi}{15}\left(k_{F}a_{s}\right)^{3}\left(\frac{T}{T_{F}}\right)^{2},\label{eq:tanFL}\end{equation}
 where the zero temperature homogeneous contact ${\cal I}_{0}$ can
be calculated using the Lee and Yang's ground state energy (up to
$(k_{F}a_{s})^{2}$) and has the form \cite{tan2}, \begin{equation}
\frac{{\cal I}_{0}}{Nk_{F}}=\frac{4}{3}\left(k_{F}a_{s}\right)^{2}\left[1+\frac{12\left(11-2\ln2\right)}{35\pi}k_{F}a_{s}\right].\end{equation}
 It is clear from Fig. 7\textit{b} that the weak-coupling Fermi-liquid
prediction works at very low temperatures only and strongly over-estimates
the contact once $T>0.2T_{F}$, although the theory describes qualitatively
the correct behavior of an increasing contact up to $T\sim T_{F}$.
Near the unitarity limit, the use of Eq. (\ref{eq:tanFL}) and hence
its support for an enhancement of the contact is certainly doubtful.

In brief, we feel that in the unitarity limit it is more reasonable
to expect a monotonically decreasing contact as the temperature increases.
It is interesting to note that, for a weaker coupling constant, $1/(k_{F}a_{s})=-1$,
all the strong-coupling theories give the same qualitative behavior
of the contact, as shown in the inset of Fig. 6. In particular, the
contact is predicted to decrease with temperature in the superfluid
phase. It is natural to assume that the same decrease would happen
as well for a unitary Fermi superfluid.

At high temperatures, we find that the NSR and $G_{0}G_{0}$ theories
predict essentially the same result, as the difference between the
two theories becomes rather small in the high-temperature regime.
However, both predictions lie systematically below the quantum virial
expansion result, indicating the inaccuracies of these theories. In
contrast, at high temperatures the self-consistent $GG$ result approaches
the second-order virial expansion prediction. None of the strong-coupling
theories can produce correctly the third-order expansion result, since
the many-body \textit{T}-matrix approach fails to account for the
three-particle scattering process.

\section{Low-$T$ trapped contact from strong coupling theories}

Let us now turn to the contact of a unitary Fermi gas in a harmonic
trap. Under the local density approximation this is given by \begin{equation}
{\cal I}_{T}=\int d{\bf r}\mathcal{C}({\bf r})=(3\pi^{2})^{1/3}\int d{\bf r[}\mathcal{C}/(\rho k_{F})]({\bf r})\rho^{4/3}({\bf r})\,,\end{equation}
where $\rho({\bf r})$ is the density distribution and $k_{F}({\bf r})=[3\pi^{2}\rho({\bf r})]^{-1/3}$.

In the unitarity limit, the local contact ${\bf [}\mathcal{C}/(\rho k_{F})]({\bf r})$
is a function of the density only, through the reduced temperature
$T/T_{F}({\bf r})$. The density may be expressed using $\rho=8/(3\sqrt{\pi})\lambda^{-3}(T_{F}/T)^{3/2}$,
so that the trapped contact is given by, \begin{equation}
{\cal I}_{T}=\frac{4}{3\lambda^{4}}\int\limits _{0}^{\infty}4\pi r^{2}dr\left(\frac{\mathcal{C}}{\rho k_{F}}\right)\left(r\right)\frac{T_{F}^{2}\left(r\right)}{T^{2}}.\end{equation}
 Here, for simplicity, we assume an isotropic trap. In the local density
approximation, it is convenient to use a local inverse fugacity $\zeta\left(r\right)=e^{-\mu(r)/k_{B}T}\equiv\zeta_{0}\exp[V_{T}\left(r\right)/k_{B}T]$
with $\zeta_{0}=e^{-\mu/k_{B}T}$. The integration over the radius
$r$ can then be converted to an integration about the inverse fugacity,
by using $r=\sqrt{(2k_{B}T/m\omega^{2})\ell(\zeta)}$ with $\ell(\zeta)\equiv\ln(\zeta/\zeta_{0}).$
In a harmonic trap, the total number of atoms is related to the Fermi
temperature by $N=(1/3)(k_{B}T_{F})^{3}/(\hbar\omega)^{3}$ and the
Fermi wave-vector is given by $k_{F}=\sqrt{2mk_{B}T_{F}/\hbar^{2}}$.
Thus, we find that, \begin{equation}
\left(\frac{{\cal I}}{Nk_{F}}\right)_{T}=\frac{32}{\pi}\frac{T^{7/2}}{T_{F}^{7/2}}\int\limits _{\zeta_{0}}^{\infty}d\zeta\frac{d\sqrt{\ell(\zeta)}}{d\zeta}\ell(\zeta)\frac{\tilde{I}\left(\zeta\right)}{\tilde{T}^{2}\left(\zeta\right)},\label{trapContact}\end{equation}
 where the inputs, the local dimensionless contact $\tilde{\mathcal{I}}\equiv{\cal C}/(\rho k_{F})$
and the local reduced temperature $\tilde{T}\equiv T/T_{F}$, are
functions of the inverse fugacity $\zeta$, and can be determined
using the homogeneous equation of state. The temperature $T/T_{F}$
in the above equation (\ref{trapContact}) is yet to be related to
the inverse fugacity $\zeta_{0}$. For this purpose, we rewrite the
number equation $N=\int d{\bf r}\rho({\bf r})$ into the form, \begin{equation}
\left(\frac{T_{F}}{T}\right)^{3}=\frac{32}{\pi}\int\limits _{\zeta_{0}}^{\infty}d\zeta\frac{d\sqrt{\ell\left(\zeta\right)}}{d\zeta}\frac{\ell\left(\zeta\right)}{\tilde{T}\left(\zeta\right)^{3/2}}.\label{trapNum}\end{equation}
 For a given $\zeta_{0}$, we use Eqs. (\ref{trapNum}) and (\ref{trapContact})
to calculate the reduced temperature $T/T_{F}$ and the trapped contact
$[{\cal I}/(Nk_{F})]_{T}$, respectively, and in turn express the
contact as a function of the reduced temperature. The detailed procedure
of numerical calculations can be found in the Appendix B.

\subsection{Trapped contact at zero temperature}

At zero temperature, the trapped contact at unitarity can be found
analytically. In this case, the density profile is known exactly,
\begin{equation}
\rho\left(r\right)=\rho_{0}\left(1-\frac{r^{2}}{R_{TF}^{2}}\right)^{3/2},\end{equation}
 where the peak density $n_{0}$ and the Thomas-Fermi $R_{TF}$ are
respectively given by, \begin{eqnarray}
\rho_{0} & = & \xi^{-3/4}\frac{(24N)^{1/2}}{3\pi^{2}}\left(\frac{\hbar}{m\omega}\right)^{-3/2},\\
R_{TF} & = & \xi^{1/4}(24N)^{1/6}\sqrt{\frac{\hbar}{m\omega}},\end{eqnarray}
 where $\xi\simeq0.41(1)$ is the universal parameter of a unitary
Fermi gas \cite{ensEoS2}. Because the temperature is zero, the local
dimensionless contact $[\mathcal{{\cal C}}/(N\rho k)]_{\hom}$ is
spatially independent, as it can only depend on a scale factor which
has already been included in the definition of this quantity. Here,
for clarity we explicitly indicate the homogeneous contact by the
suffix {}``$\hom$''. 

Thus, the trapped zero-temperature contact is given by, \begin{eqnarray}
{\cal I}_{T} & = & (3\pi^{2})^{1/3}\left(\frac{\mathcal{C}}{\rho k_{F}}\right)_{\hom}\int d{\bf r}\rho^{4/3}({\bf r}),\\
 & = & \frac{32\pi}{105}(3\pi^{2})^{1/3}\left(\frac{\mathcal{C}}{\rho k_{F}}\right)_{\hom}\rho_{0}^{4/3}R_{TF}^{3}.\end{eqnarray}
 Using $k_{F}=\sqrt{2mk_{B}T_{F}/\hbar^{2}}$ in harmonic traps with
$k_{B}T_{F}=(3N)^{1/3}\hbar\omega$, we find then, \begin{equation}
\left(\frac{{\cal I}}{Nk_{F}}\right)_{T}=\frac{256}{105\pi}\xi^{-1/4}\left(\frac{\mathcal{C}}{\rho k_{F}}\right)_{\hom}=\frac{512\zeta}{175\xi^{1/4}},\end{equation}
 where the two universal parameters $\xi$ and $\zeta$ are related
to the ground state energy of an interacting Fermi gas near unitarity
\cite{tan2}, $E/(N\epsilon_{F})=(3/5)[\xi-\zeta/(k_{F}a_{s})+\cdots]$.

\begin{table}[h]
 \begin{tabular}{cc}
 & \tabularnewline
\hline
\hline 
$\left({\cal I}/(Nk_{F})\right)_{T}$  & Methods \tabularnewline
\hline 
$3.00\pm0.12$  & Swinburne Bragg spectroscopy (Ref. \cite{swinTan}) \tabularnewline
$3.40\pm0.18$  & ENS equation of state data (Ref. \cite{ensEoS2}) \tabularnewline
\hline 
$3.26$  & GPF (present work) \tabularnewline
$3.03$  & $GG$ (Ref. \cite{haussmann}) \tabularnewline
$3.05$  & $G_{0}G_{0}$ (Ref. \cite{palestini}) \tabularnewline
\hline
\end{tabular}

\caption{Zero temperature trapped contact in the unitarity limit. We compare
different results obtained from the three strong-coupling theories
with the experimental measurements at Swinburne \cite{swinTan} and
ENS \cite{ensEoS2}.}

\end{table}

Table II lists the zero-temperature contact of a unitary Fermi in
harmonic traps. The theoretical prediction from different strong-coupling
theories are compared with the experimental measurements at the Swinburne
University of Technology, Melbourne \cite{swinTan} and at ENS, Paris
\cite{ensEoS2}. We note that, the Swinburne data point is determined
from spin-antiparallel static structure factor measured by Bragg spectroscopy.
The lowest temperature in the Bragg measurement is about $0.10\pm0.02T_{F}$,
which could explain in part the slightly lower value obtained for
the contact. For the ENS data point, we have used $\xi=0.41(1)$ and
$\zeta=0.93(5)$, as determined from the experimental data for ground
state energy. The temperature of this measurement was below the limits
of the thermometry used, but we can estimate this as $0.05\pm0.05T_{F}$
from comparisons with related experimental measurements.

\subsection{Universal trapped contact from strong-coupling theories}

\begin{figure}[htp]

\begin{centering}
\includegraphics[clip,width=0.45\textwidth]{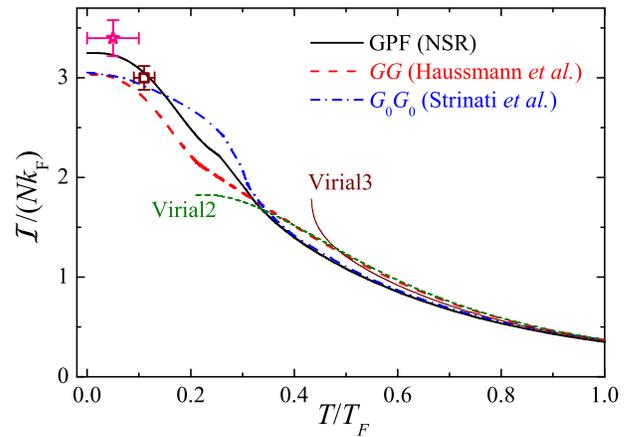} 
\par\end{centering}

\caption{(Color online) Universal contact of a trapped Fermi gas in the unitarity
limit, calculated from different strong-coupling theories as well
as from the quantum virial expansion method. The symbols indicate
the measurements at ENS, Paris (star) \cite{ensEoS2} and at the Swinburne
University of Technology, Melbourne (square) \cite{swinTan}.}

\label{fig8} 
\end{figure}

Figure 8 reports the temperature dependence of the contact of a trapped
Fermi gas at unitarity, obtained from different strong-coupling theories.
We compare the theoretical results with two experimental data points
measured at low temperatures, while at high temperatures, the results
are contrasted with the accurate quantum virial expansion prediction.

At low temperatures, $T<0.3T_{F}$, there is a sizable discrepancy
between the predictions from different strong-coupling theories. Nevertheless,
all the predictions are now in qualitative agreement, as the maximum
in the homogeneous contact found previously in the $G_{0}G_{0}$ theory
is washed out by the trap averaging \cite{palestini}. At high temperatures,
the difference between different theories become much smaller. This
is in line with our observation for the $n$-th contact coefficient,
which receives a factor of $n^{3/2}$ reduction in harmonic traps.

The trapped contact of a unitary Fermi gas can be measured readily
in experiment, by extending the existing low-temperature measurements
of either the momentum distribution \cite{jilaTan} or Bragg spectroscopy
of a unitary Fermi gas \cite{swinTan} to the high-temperature regime.
The precise determination of the temperature in the unitarity limit
is a challenging problem \cite{dukeEoS2}. However, this can be overcome
by using the known equation of state in the unitarity limit \cite{ensEoS1,hldcmp06,hldcmp10}.
We anticipate that the quantitative discrepancy between different
strong-coupling theories shown in Fig. 8 will be clarified by future
experiments.

\section{Experimental measurement of the homogeneous contact at unitarity}

It would be very useful to determine the contact of a homogeneous
unitary Fermi gas, without the complication caused by the trap averaging.
The result would provide a unique opportunity to test quantum many-body
theories of strongly interacting Fermi gases that we have discussed
above. However, an {\em in-situ} measurement is generally required,
which is difficult to setup experimentally. The determination of the
universal homogeneous contact could be achieved as follows, by:
\begin{itemize}
\item Measurement of homogeneous equation of state near unitarity. The Salomon
group at ENS, Paris has already demonstrated the accurate measurements
of uniform equation of state either in the unitarity limit \cite{ensEoS1}
or at zero temperature \cite{ensEoS2}. The measurement could be extended
straightforwardly to the general case with arbitrary coupling constant
and temperature, say, for example, the entropy $S(T/T_{F},1/k_{F}a)$.
The homogeneous contact can then be determined directly using Eq.
(\ref{dIdT}).
\item Measurement of the tail of the spatially-resolved RF spectrum at unitarity.
Tomographic RF spectroscopy was recently developed at MIT to measure
the local pairing gap of a unitary Fermi gas of $^{6}$Li atoms at
finite temperatures \cite{mitRF}. Ideally, the local universal contact
could be retrieved from the tail of RF-spectrum that obeys a $\omega^{-3/2}$
power-law. However, the analysis of the tail structure is complicated
for $^{6}$Li atoms due to a strong final state effect. It would be
desirable to develop a spatially-resolved RF-spectroscopy for $^{40}$K
atoms, for which the final state effect may be negligible.
\item Measurement of the spatially-resolved Bragg scattering spectrum at
unitarity \cite{swinTan,swinBragg}. Thus, one obtains the local spin-antiparallel
static structure factor at large momentum $q$, which at unitarity
satisfies the asymptotic behavior of $S_{\uparrow\downarrow}(q)=[{\cal I}/(Nk_{F})]/(4q/k_{F})$.
The local universal contact is then determined. 
\end{itemize}
\begin{figure}[htp]

\begin{centering}
\includegraphics[clip,width=0.45\textwidth]{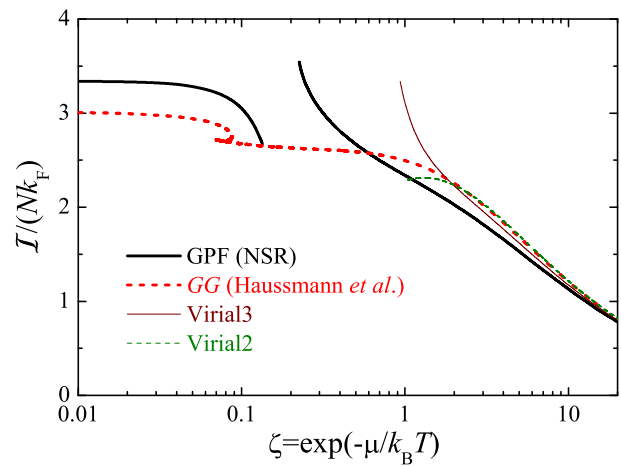} 
\par\end{centering}

\caption{(Color online) The universal contact of a homogeneous unitary Fermi
gas as a function of the inverse fugacity, calculated using the Gaussian
pair fluctuation theory, self-consistent $GG$ theory and quantum
virial expansion.}

\label{fig9} 
\end{figure}

All of these measurements are based on the local density approximation.
The local contact is then to be determined as a function of the local
chemical potential $\mu(r)$ or local inverse fugacity $\zeta(r)=\exp(-\mu(r)/k_{B}T)$.
In Fig. 9, we show the inverse fugacity dependence of the universal
contact of a unitary Fermi gas, to be confronted with future experiments.

\section{Conclusions}

To summarize, we have presented a comparative theoretical study of
the universal contact for a strongly interacting Fermi gas at unitarity.
The contact at high temperature has been accurately determined by
using a quantum virial expansion method, while at low temperatures,
we have employed different strong-coupling theories to try to estimate
the temperature dependence of the contact.

The temperature dependence of the contact near the critical temperature,
however, remains unresolved, since one of the strong-coupling theories,
the $G_{0}G_{0}$ theory, predicts an enhancement or a maximum near
criticality, which is not observed in the other two strong-coupling
theories. We believe that this enhancement could be due to the inaccuracy
of the $G_{0}G_{0}$ approach and conjecture that the universal contact
should decrease monotonically with increasing temperature, except
possibly at very low temperatures $T\ll T_{c}$. 

As there is no solid justification for any of these strong-coupling
theories, our conjecture should be examined critically by future experiments
or advanced quantum Monte Carlo simulations. Therefore, we have proposed
several experimental ways to determine the homogeneous contact of
a unitary Fermi gas, using measurements of equation of state, tomographic
RF-spectroscopy, and spatially-resolved Bragg spectroscopy. All of
these measurements are within reach of current experimental techniques.

We note finally that fermionic universality and the detailed behavior
of a strongly interacting Fermi gas near the normal-superfluid transition
is of great interest. Tan's contact provides an entirely new means
to characterize the universal properties and phases of strongly interacting
fermions, in addition to the equation of state. Our comprehensive
study of the universal contact, built on the most recent theoretical
methods, should provide useful insights for future research.

\textit{Note added. }--- Since the submission of our manuscript, the
finite-temperature contact of a trapped unitary Fermi gas was measured
at Swinburne \cite{finiteTSwinExpt}. The experimental result agrees
fairly well the theoretical predictions in Fig. 8. A quantum Monte
Carlo simulation of the finite-temperature contact of a homogeneous
unitary Fermi gas was also carried out by Drut, Lähde and Ten \cite{finiteTQMC}.
\begin{acknowledgments}
We thank S. Tan, D. S. Jin, C. Vale, E. Kuhnle, C. Salomon and F.
Werner for discussions and acknowledge R. Haussmann, T. Enss, and
F. Palestini for providing us with their theoretical predictions for
the universal contact. This work is supported by the ARC Centre of
Excellence, ARC Discovery Projects No. DP0984522 and No. DP0984637,
and NSFC Grant No. 10774190. Any correspondence should be addressed
to PDD at pdrummond@swin.edu.au.
\end{acknowledgments}
\appendix

\section{Contact from the vertex function}

In the many-body theory, the contact is given by the vertex function
in the limits of short distance and time.

\begin{figure}[htp]
\begin{centering}
\includegraphics[clip,width=0.35\textwidth]{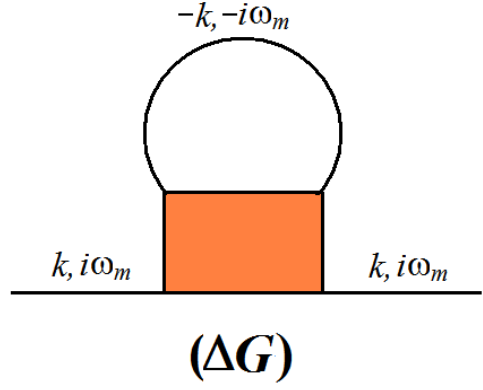} 
\par\end{centering}

\caption{Diagram contribution to $\Delta G(x,\tau)=G(x,\tau)-G^{(1)}(x,\tau)$
at short distance and time. The shadow region indicates the vertex
function at $x=0$ and $\tau=0^{-}$.}

\label{figapp} 
\end{figure}

For a normal Fermi gas, this can be understood by calculating the
large-frequency tail of the momentum distribution $\rho_{\sigma}\left(k\right)$.
For this purpose, we consider the diagram contribution to $\Delta G(x,\tau=0^{-})=G(x,\tau=0^{-})-G^{(1)}(x,\tau=0^{-})$,
as shown in Fig. 10, where $G^{(1)}(x,\tau)$ is the finite-temperature
Green function of an ideal Fermi gas. In the limit of large momentum
$k\rightarrow\infty$ or short distance $x\rightarrow0$, we may assume
that the vertex function $\Gamma(x,\tau=0^{-})$ is a smooth function
and therefore set $\Gamma(x,\tau=0^{-})\simeq\Gamma(0,0^{-})$. According
to Fig. 10, the finite temperature Green function in the momentum
space is then given by ($\mu=0$), \begin{equation}
\Delta G\left(k,i\omega_{m}\right)\simeq-\Gamma(0,0^{-})\frac{1}{\left[i\omega_{m}-\epsilon_{k}\right]^{2}}\frac{1}{\left(i\omega_{m}+\epsilon_{k}\right)},\end{equation}
 where $\omega_{m}=(2m+1)\pi k_{B}T$ is the fermionic Matasubara
frequency and $\epsilon_{k}\equiv\hbar^{2}k^{2}/(2m)$. The momentum
distribution $\Delta\rho_{\sigma}\left(k\right)=k_{B}T\sum_{m}\Delta G\left(k,i\omega_{m}\right)$
takes the form, \begin{eqnarray}
\Delta\rho_{\sigma}\left(k\right) & \simeq & -\Gamma(0,0^{-})\frac{1}{\left[e^{-\epsilon_{k}/(k_{B}T)}+1\right]}\frac{1}{4\epsilon_{k}^{2}},\\
 & \simeq & -\frac{m^{2}}{\hbar^{4}}\Gamma(0,0^{-})\frac{1}{k^{4}}.\end{eqnarray}
 Thus, the contact is given by, ${\cal I}=-m^{2}\Gamma(0,0^{-})/\hbar^{4}$
or \begin{equation}
\frac{{\cal I}}{Nk_{F}}=-\frac{3\pi^{2}}{4}\frac{\Gamma\left({\bf x}=0,\tau=0^{-}\right)}{\epsilon_{F}^{2}}.\end{equation}

In the superfluid phase, we should explicitly introduce the pairing
gap and use a 2 by 2 matrix for the Green function and vertex function.
This leads to Eq. (\ref{eq:contactVertex}).

\section{Calculation of the integrals in the local density approximation}

The two integrand functions in Eqs. (\ref{trapContact}) and (\ref{trapNum}),
which involve the local reduced temperature, are ill-behaved at low
temperatures. Therefore, we regularize the reduced temperature by
its non-interacting value and define, \begin{equation}
t\left(\zeta\right)\equiv\frac{\tilde{T}\left(\zeta\right)}{\tilde{T}_{IG}\left(\zeta\right)},\end{equation}
 where $\tilde{T}_{IG}\left(\zeta\right)$ is the reduced temperature
of an ideal, non-interacting Fermi gas, defined by: \begin{eqnarray}
\tilde{T}_{IG}\left(\zeta\right) & = & \left[f\left(\zeta\right)\right]^{-2/3},\end{eqnarray}
where: \begin{equation}
f\left(\zeta\right)=\frac{3}{2}\int\limits _{0}^{\infty}dt\sqrt{t}\frac{\zeta^{-1}e^{-t}}{1+\zeta^{-1}e^{-t}}.\end{equation}
The coupled equations then take the form,

\begin{eqnarray}
\frac{{\cal I}_{T}}{Nk_{F}} & = & \frac{32}{\pi}\frac{T^{7/2}}{T_{F}^{7/2}}\int\limits _{\zeta_{0}}^{\infty}d\zeta\frac{d\sqrt{\ell\left(\zeta\right)}}{d\zeta}\ell\left(\zeta\right)f^{4/3}\left(\zeta\right)\frac{\tilde{I}\left(\zeta\right)}{t^{2}\left(\zeta\right)},\label{trapContactAA}\\
\frac{T_{F}^{3}}{T^{3}} & = & \frac{32}{\pi}\int\limits _{\zeta_{0}}^{\infty}d\zeta\frac{d\sqrt{\ell\left(\zeta\right)}}{d\zeta}\ell\left(\zeta\right)f\left(\zeta\right)\left[t\left(\zeta\right)\right]^{-3/2},\label{trapNumAA}\end{eqnarray}
 where $\ell\left(\zeta\right)\equiv\ln(\zeta/\zeta_{0})$, and $\tilde{I}\left(\zeta\right)/t^{2}\left(\zeta\right)$
and $[t\left(\zeta\right)]^{-3/2}$ are now smooth functions of $\zeta$.

To do the integration, for instance, for the integral in the number
equation (\ref{trapNumAA}), \begin{equation}
A=\int\limits _{\zeta_{0}}^{\infty}d\zeta\frac{d\sqrt{\ell\left(\zeta\right)}}{d\zeta}\ell\left(\zeta\right)f\left(\zeta\right)\left[t\left(\zeta\right)\right]^{-3/2},\end{equation}
 we use the following discretized version, \begin{equation}
A=\sum_{\zeta_{N-1}\geqslant\zeta_{i}\geqslant\zeta_{0}}I\left(\zeta_{i}\right)+I_{\infty},\end{equation}
 where the two types of integrals $I\left(\zeta_{i}\right)$ and $I_{\infty}$
are defined by, \begin{eqnarray}
A\left(\zeta_{i}\right) & = & \int\limits _{\zeta_{i}}^{\zeta_{i+1}}d\zeta\frac{d\sqrt{\ell\left(\zeta\right)}}{d\zeta}\ell\left(\zeta\right)f\left(\zeta\right)\left[t\left(\zeta\right)\right]^{-3/2},\\
A_{\infty} & = & \int\limits _{\zeta_{N}}^{\infty}d\zeta\frac{d\sqrt{\ell\left(\zeta\right)}}{d\zeta}\ell\left(\zeta\right)f\left(\zeta\right)\left[t\left(\zeta\right)\right]^{-3/2},\end{eqnarray}
 respectively. Here, $\{\zeta_{i}\}$ ($i=1,...,N$) are the set of
points where $t\left(\zeta\right)$ has values. We shall let $\zeta_{0}$
run over all the points of $\{\zeta_{i}\}$. To calculate $A\left(\zeta_{i}\right)$,
because of the smoothness of $t\left(\zeta\right)$ we approximate
it by, \begin{eqnarray}
A\left(\zeta_{i}\right) & \approx & \overline{\left[t\left(\zeta\right)\right]^{-3/2}}\int\limits _{\zeta_{i}}^{\zeta_{i+1}}d\zeta\frac{d\sqrt{\ell\left(\zeta\right)}}{d\zeta}\ell\left(\zeta\right)f\left(\zeta\right),\\
 & = & \overline{\left[t\left(\zeta\right)\right]^{-3/2}}\int\limits _{\sqrt{\ell\left(\zeta_{i}\right)}}^{\sqrt{\ell\left(\zeta_{i+1}\right)}}dxx^{2}f\left(\zeta_{0}e^{x^{2}}\right),\end{eqnarray}
 where $\overline{\left[t\left(\zeta\right)\right]^{-3/2}}\equiv[t^{-3/2}\left(\zeta_{i}\right)+t^{-3/2}\left(\zeta_{i+1}\right)]/2$,
and in the second line we have changed to a new variable $x\equiv\sqrt{\ell\left(\zeta\right)}$.
For the integral $A_{\infty}$, where $\zeta_{N}$ is sufficiently
large, we may use the virial expansion result, \textit{i.e.}, \begin{equation}
\left[t_{\text{ve}}\left(\zeta\right)\right]^{-3/2}=1+\frac{3\sqrt{\pi}}{4}\frac{\Delta b_{2,\infty}\zeta^{-2}+\Delta b_{3,\infty}\zeta^{-3}}{f\left(\zeta\right)},\end{equation}
 with $\Delta b_{2,\infty}=1/\sqrt{2}$ and $\Delta b_{3,\infty}=-0.35510298$.
Thus, we have, \begin{equation}
A_{\infty}=\int\limits _{\sqrt{\ell\left(\zeta_{N}\right)}}^{\infty}dxf\left(\zeta_{0}e^{x^{2}}\right)\left[t_{\text{ve}}\left(\zeta_{0}e^{x^{2}}\right)\right]^{-3/2}.\end{equation}
 The integral in the trapped contact equation (\ref{trapContactAA})
can be done in a similar way. We note that at high temperatures, \begin{equation}
\left[\frac{\tilde{I}\left(\zeta\right)}{t^{2}\left(\zeta\right)}\right]_{\text{ve}}=\frac{3\pi^{2}\left[c_{2,\infty}\zeta^{-2}+c_{3,\infty}\zeta^{-3}+\cdots\right]}{\left[f\left(\zeta\right)\right]^{4/3}},\end{equation}
 with $c_{2,\infty}=1/\pi$ and $c_{3,\infty}=-0.141$.

\end{document}